**5DSTEM of Liquid-Phase Nanocrystal Growth Bridges Ensemble Kinetics and Nanoscale Dynamics**

Authors: Serin Lee[1,*], Colin Ophus[1,*], Jennifer Dionne[1,2,*]
1: Department of Materials Science and Engineering, Stanford University, Stanford, CA, USA.
2: Department of Radiology, Stanford University School of Medicine, Stanford, CA, USA.
Corresponding authors (*): serinl@stanford.edu, cophus@stanford.edu, jdionne@stanford.edu

**Abstract**

Resolving how ensemble growth kinetics emerge from particle- and grain-level dynamics is essential for predictive control of nanocrystal synthesis. Here, time-resolved four-dimensional scanning transmission electron microscopy (4DSTEM), extended into five dimensions (5DSTEM) through continuous acquisition, is used to simultaneously resolve morphology, crystallographic orientation, and lattice strain during Au nanocrystal growth in aqueous $HAuCl_4$. Although ensemble growth follows surface-reaction-limited kinetics, individual nanoparticles exhibit distinct pathways, including continuous growth and discrete coalescence through oriented attachment. These pathways display characteristic orientation dynamics, with stable or gradually selected orientations during continuous growth and abrupt reconfiguration followed by alignment during coalescence. At the ensemble level, the out-of-plane orientation distribution is established early and remains stable, whereas in-plane orientations remain broadly distributed without global alignment. Tensile strain develops progressively from particle surfaces and interfaces, directly linking lattice distortion with evolving morphology and crystallographic structure. By connecting ensemble kinetics with particle- and grain-resolved structural evolution, 5DSTEM provides a general framework for uncovering heterogeneous growth mechanisms and for relating synthesis pathways to structure-dependent properties in functional nanomaterials.

**Table of Contents**

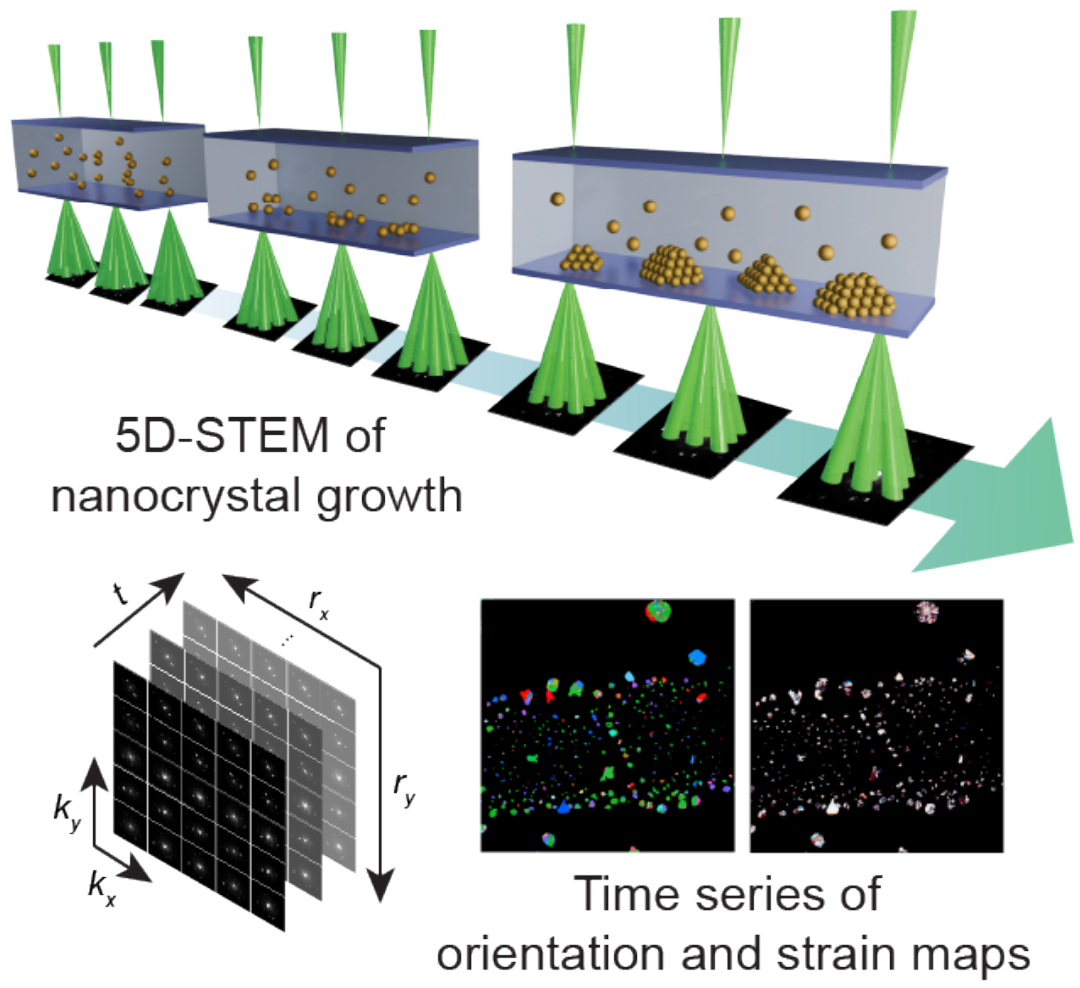

5DSTEM of nanocrystal growth connecting diffraction data over time and space to the time series of orientation and strain maps.

## Introduction

Understanding nanoparticle nucleation and growth is central to controlling nanocrystal structure and morphology, which ultimately governs performance in applications spanning catalysis, electronics, and photonics. Liquid-cell transmission electron microscopy (TEM) has emerged as a powerful platform for directly visualizing these processes in real time.[1–16] By leveraging electron-beam–induced radiolysis, liquid cell TEM enables the generation of reactive species that drive reduction reactions, effectively creating a tunable chemical environment that can mimic aspects of solution-phase synthesis [17–22]. The underlying chemistry is governed by electron–liquid interactions via radiolysis, in which energy deposition leads to the formation of solvated electrons, radicals, and molecular species, whose diffusion and reactions define the evolving local chemical environment [17,23–30]. Due to its unique ability to replicate nanocrystal synthesis conditions, liquid-cell TEM has revealed non-classical nucleation and growth pathways [19,28,29,31–33]. Recent examples of nanocrystal synthesis in liquid cell TEM include the synthesis of metastable hexagonal close-packed palladium hydride by exploiting the interplay among precursor concentrations [34] and plasmon-mediated deposition of Ag onto Au nanorods [35].

However, most in situ S/TEM studies have focused on ensemble kinetics, tracking the temporal evolution of particle-size distributions across nanocrystals and fitting them to classical models such as Lifshitz–Slyozov–Wagner (LSW) theory to reveal mechanistic pathways of nucleation and growth [20,28,30,31,36,37]. While such approaches provide valuable averaged descriptors of growth behavior, there is a need for analysis of the intrinsic heterogeneity of nanoparticle formation, in which multiple particles and grains evolve concurrently under shared but spatially varying chemical conditions. In practice, growth proceeds through multigrain intermediates and particle–particle interactions via orientation evolution that cannot be fully captured by ensemble-averaged measurements alone. [5,38–40] Establishing a direct connection between ensemble observables and per-particle or per-grain dynamics, therefore, remains a critical challenge for developing a predictive understanding of nanoscale growth pathways.

A particularly underexplored consequence of these heterogeneous growth pathways is the development of orientation and the following internal lattice strain. In catalytically active materials, such strain modifies the electronic structure of surface atoms by shifting the d-band center, thereby altering the binding energies of adsorbates and reaction intermediates, which can dramatically influence activity and selectivity [41–46]. Yet because strain accumulates dynamically and heterogeneously during growth, ensemble-averaged or post-synthesis structural characterization cannot capture how it originates or evolves. Connecting the dynamic structural pathways of nanocrystal formation to the function and properties requires a technique capable of resolving orientation, strain, and morphology simultaneously, across both individual particles and the broader ensemble, in real time.

Here, we introduce time-resolved four-dimensional scanning transmission electron microscopy (4DSTEM), extended to five dimensions (5DSTEM), as a framework for bridging ensemble kinetics and nanoscale structural dynamics. In conventional 4DSTEM, the electron probe is rastered while diffraction patterns are recorded at each scan position [47–52]. In our previous work, we discussed *in situ* 4DSTEM, or 5DSTEM, in which these datasets are continuously acquired over time [53], and developed a computational workflow for their efficient analysis.[54] Building on these foundations, we provide the first quantitative demonstration of 5DSTEM that simultaneously resolves growth kinetics, crystallographic orientation, and strain throughout a dynamic nanocrystal growth process. Specifically, we quantify ensemble kinetics across individual nanoparticles, distinguish particle-level growth pathways, and statistically map the evolution of orientation and strain across the nanocrystal ensemble. This approach directly connects ensemble growth laws with the heterogeneous particle- and grain-level structural dynamics from which they emerge, addressing a key capability previously identified for applying 5DSTEM to dynamic catalytic and energy materials [53].

Using Au nanocrystal growth in an aqueous $HAuCl_4$ solution as a model system, we achieve spatially and temporally resolved mapping of orientation and strain, which is inaccessible from morphology or size-based analysis alone, enabling direct tracking of grain-specific structural dynamics. We show that, while ensemble growth follows surface-reaction-limited kinetics, individual nanoparticles evolve through heterogeneous pathways, including continuous growth and discrete coalescence events via oriented attachment. These distinct growth modes exhibit different signatures in orientation evolution, with stable or gradually selected orientations in continuous growth and abrupt reconfiguration followed by alignment during coalescence. At the ensemble level, we further identify an anisotropic orientation evolution, in which the out-of-plane orientation distribution is established early and remains stable, whereas in-plane orientations remain broadly distributed without global alignment. We also observe the development of tensile strain originating near particle surfaces and interfaces, revealing how lattice distortion emerges and evolves alongside changes in particle morphology and orientation. Our results establish 5DSTEM as a powerful approach for directly linking ensemble growth behavior to the nanoscale mechanisms that govern nanocrystal formation, providing a pathway toward predictive control of dynamic materials synthesis across a broad range of solution-phase and catalytic systems.

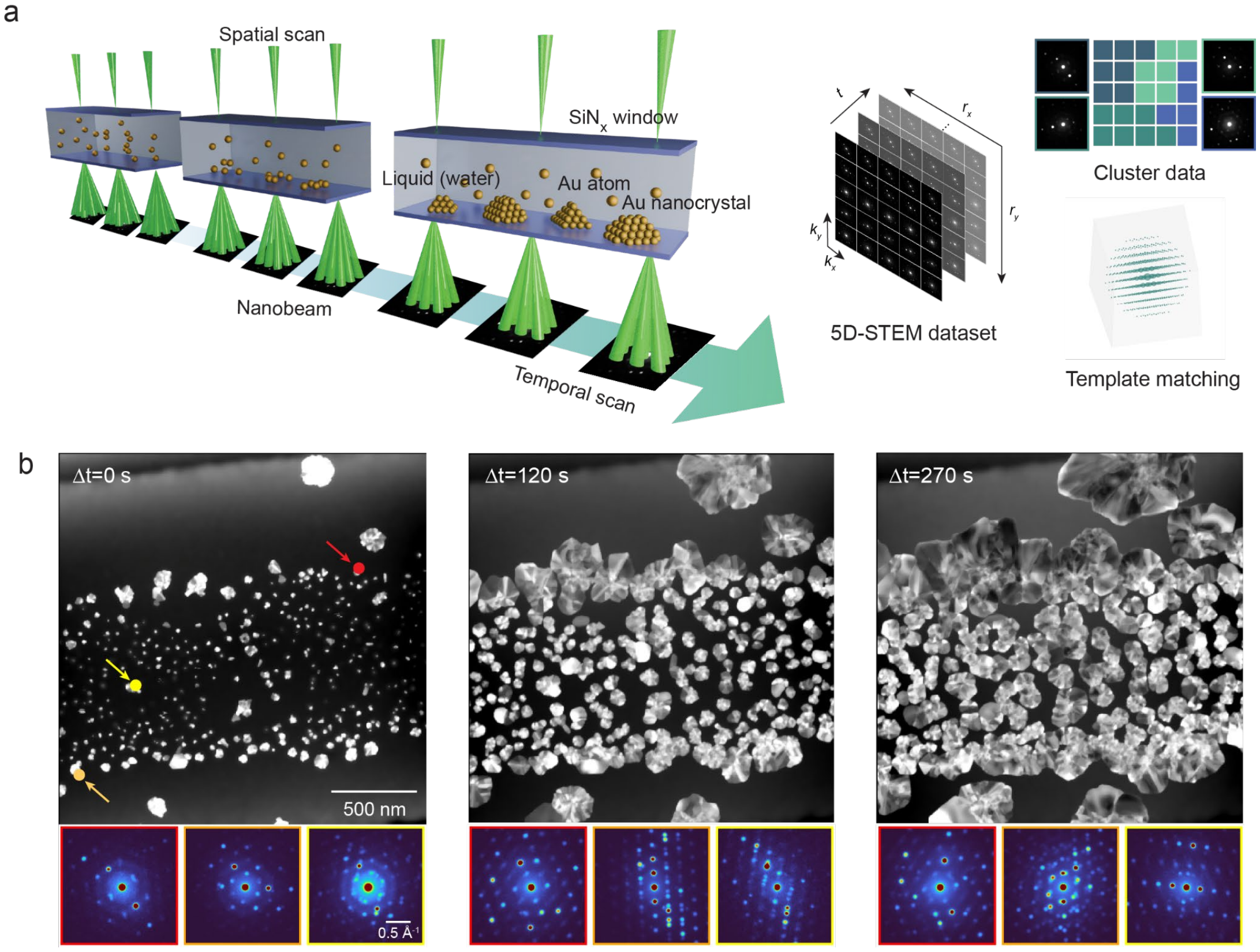

Figure 1. 5DSTEM experiment on the Au nanocrystal growth dynamics. (a) Schematic representation of the experimental workflow. (b) Representative virtual annular dark-field images and corresponding nanobeam diffraction patterns from the marked regions.

## Results and Discussion

We first illustrate the workflow of the 5DSTEM experiment to probe nanocrystal growth dynamics in real and reciprocal space, as shown in Figure 1. During liquid-cell TEM imaging, a focused electron probe is rastered across the sample in real space, while the scan is repeated sequentially over time as nanoparticles evolve. At each probe position, a nanobeam electron diffraction pattern is recorded, enabling simultaneous acquisition of spatially resolved structural information. This approach generates a five-dimensional dataset (2D real space × 2D reciprocal space × time), capturing the evolution of nanocrystal structure during growth. To extract physically meaningful information from this high-dimensional dataset, we employ a two-step analysis workflow. First, diffraction patterns are grouped using an unsupervised clustering approach based on local similarity, following our previous work [54], which enhances signal-to-noise and identifies regions with similar crystallographic features. Second, the clustered diffraction patterns are analyzed using template matching within the automated crystal orientation mapping (ACOM) framework implemented in py4DSTEM [55,56], allowing spatially resolved determination of crystal orientation and strain. The experiments are performed by introducing an aqueous 1 mM $HAuCl_4$ solution into a liquid-cell holder (Insight Chips, flow holder) with an 80 nm-thick channel. Under electron-beam irradiation, with a low electron dose rate of 1.14 e-/ $Å^2$ s,[28,33,36] radiolysis-driven reduction induces nucleation and growth of Au nanocrystals (See Methods for details).

Figure 1b presents representative virtual annular dark-field (ADF) images at selected time points ($\Delta t$ = 0 s, 120 s, and 270 s), illustrating the evolution from dispersed nuclei to densely packed, faceted nanocrystals (Full time series in Supporting Movie S1). Corresponding nanobeam diffraction patterns from marked regions out of 512 x 512 diffraction patterns are shown on the right side of the ADF images. These diffraction patterns reveal the crystallinity and orientation of individual domains, demonstrating the capability of 5DSSTEM to directly correlate real-space morphology with reciprocal-space structural information during dynamic growth.

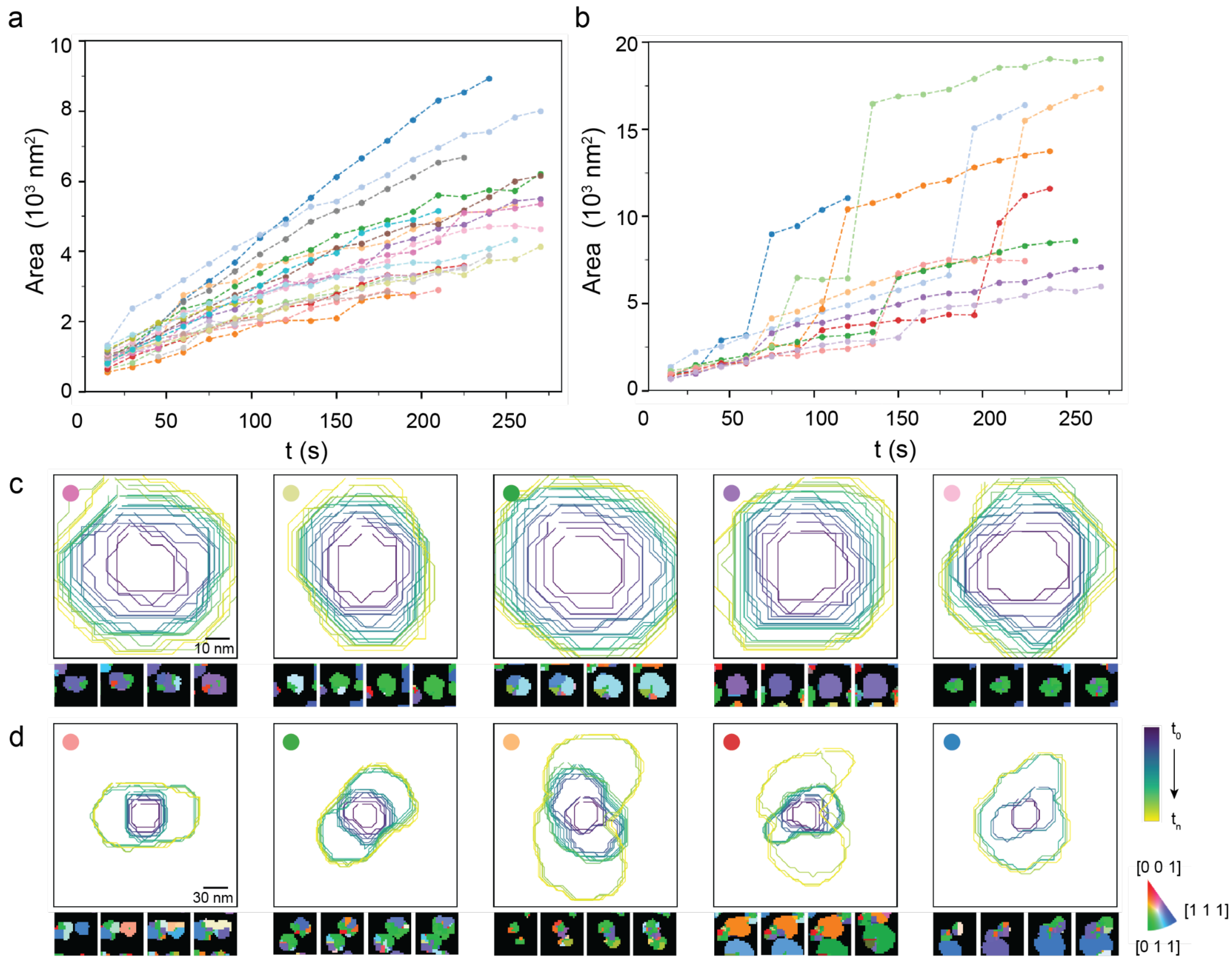

Figure 2. Ensemble kinetics and particle-resolved growth pathways. (a,b) Time-resolved projected area evolution of individual nanoparticles extracted from virtual ADF image series using automated segmentation and tracking. (c,d) Centroid-aligned contour evolution of representative nanoparticles corresponding to (a) and (b), respectively, with associated orientation maps with a time interval of 15 s.

Building on the 5DSTEM framework established in Figure 1, we next quantify ensemble growth kinetics by tracking the projected area of individual nanoparticles over time. Nanoparticle contours and corresponding areas are extracted from each ADF frame using automated segmentation based on the Segment Anything Model (SAM) (see Methods). Figure 2a shows representative particles (n = 20) that grow continuously without merging events. These particles exhibit a monotonic, near-linear increase in projected area with time, with a mean $R^2$ of 0.97 ( Supporting Figure S1a and Supporting Table 1, Supporting Information). This behavior is consistent with surface-reaction-limited growth observed under low-electron current [36], where the particle radius follows $r \propto t^{1/2}$, corresponding to a linear scaling of area with time as predicted by Lifshitz–Slyozov–Wagner (LSW) theory under reaction-limited conditions [20,30,36]. In contrast, Figure 2b shows representative particles (n = 10) undergoing migration and coalescence. Their area evolution displays three distinct stages: a linear growth regime before coalescence, an abrupt jump during coalescence, and a second linear regime afterward, with a mean $R^2$ of 0.95 for every linear segment (Supporting Figure S1b and Supporting Table 2, Supporting Information). This is consistent with the persistence of surface-reaction-limited kinetics before and after coalescence.  The growth rate generally decreased after coalescence relative to its pre-coalescence value, except for particles affected by fitting artifacts arising from discontinuous tracking between time points, showing that coalescence did not measurably accelerate subsequent growth. These are reproducible across independent liquid-cell experiments, with an additional dataset acquired under the same experimental conditions exhibiting consistent surface-reaction-limited growth behavior (Supporting Figure S2, Supporting Information). To directly connect these kinetic signatures with nanoscale structural evolution, Figures 2c and 2d show centroid-aligned contour evolution of representative particles, with corresponding orientation maps displayed below each sequence. The colored markers in the contour plots correspond to the trajectories shown in Figures 2a and 2b, and the orientation map during growth is shown below each contour map. For particles undergoing continuous growth (Figure 2c), the contours reveal faceted, anisotropic expansion with largely preserved shape symmetry. The corresponding orientation maps indicate either stable single-crystal growth or the gradual dominance of a single orientation from an initially multigrain structure. In contrast, particles undergoing coalescence (Figure 2d) exhibit abrupt morphological changes, followed by continued faceted growth. The orientation maps capture the approach and interaction of grains with distinct crystallographic orientations, followed by alignment and merging into a unified lattice. In particular, the emergence and subsequent growth of a common orientation after coalescence is consistent with an oriented-attachment mechanism [38,39].

These results demonstrate that nanocrystal growth follows surface-reaction-limited kinetics at the ensemble level, while particle-level pathways include both continuous growth and discrete coalescence events. Importantly, coalescence shows as stepwise increases in particle size without altering the underlying growth mechanism. At the same time, the orientation evolution reveals distinct signatures for each growth mode, where particles undergoing continuous growth maintain a stable crystallographic orientation or exhibit gradual selection of a dominant grain, whereas particles undergoing coalescence display abrupt reconfiguration of orientations, followed by alignment into a common lattice consistent with oriented attachment. This direct correlation between kinetic behavior and orientation dynamics demonstrates how ensemble growth laws emerge from heterogeneous, orientation-dependent nanoscale processes.

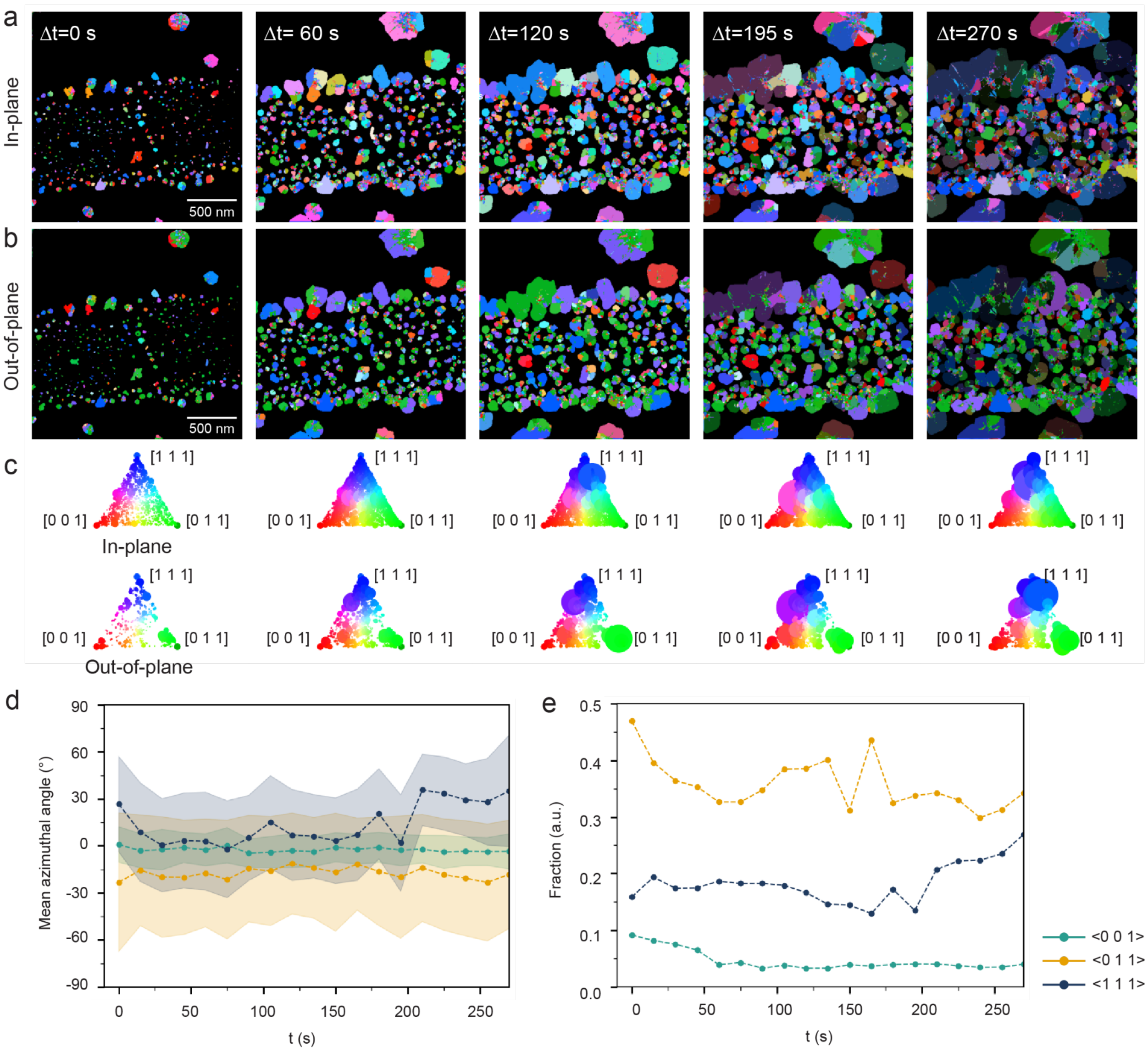

Figure 3. Temporal evolution of crystallographic orientation during Au nanocrystal growth. (a,b) In-plane (a) and out-of-plane (b) orientation maps at selected time points. (c) Corresponding orientation distributions, where marker size reflects relative population. (d) Mean azimuthal angle and the standard deviation of the in-plane orientation. (e) Fraction of dominant out-of-plane orientation.

Extending the particle-resolved analysis, we next examine how crystallographic orientations evolve at the ensemble level during growth. Figures 3a and 3b show the temporal evolution of in-plane and out-of-plane orientation maps, respectively (full time series in Supporting Movie S2), providing a spatially resolved view of orientation across the nanoparticle ensemble. To quantify these trends, Figure 3c presents the corresponding orientation distributions, where the marker size reflects the relative population of pixels assigned to each orientation. These distributions reveal how crystallographic orientations are populated and evolve over time across the entire field of view.

The in-plane orientation is analyzed by measuring the azimuthal rotation of the crystal lattice about the zone-axis for each pixel. After restricting the analysis to pixels associated with selected out-of-plane families and accounting for crystallographic symmetry and circular statistics, the in-plane distributions exhibit stable mean azimuthal angles throughout the time evolution (Figure 3d). The absence of systematic shifts in the mean indicates that no global in-plane alignment emerges during growth. Instead, each orientation family maintains a finite angular spread, reflecting rotational heterogeneity within the plane perpendicular to the zone axis.

The out-of-plane orientation is analyzed by quantifying the fraction of pixels aligned with specific zone-axis families (Figure 3e) within a threshold. We find that the ensemble is dominated by <011> and <111>-like orientations, while the [001] like population remains comparatively minor. These fractions remain largely constant over time, indicating that the out-of-plane orientation distribution is established early during nucleation and persists throughout growth. The out-of-plane fractions represent the fraction of valid pixels whose beam-aligned crystallographic direction lies within 15° of the nearest ⟨001⟩, ⟨011⟩, or ⟨111⟩ zone-axis family. Orientations more than 15° from all three families were left unclassified, ranging from 0.28 to 0.50 across time frames. We also note that variations in both in-plane and out-of-plane orientation metrics may partially arise from limitations of the orientation mapping and clustering framework, which assumes a single dominant diffraction signal per probe position. In regions with mixed diffraction contributions, such as overlapping grains or grain boundaries, ambiguity in diffraction features can increase variability in the extracted orientation distributions [57]. Nonetheless, the temporal stability of these populations suggests that the out-of-plane orientation distribution was already established at the earliest measured time point and remained stable during growth.

Together, these results reveal an anisotropic evolution of orientation during nanocrystal growth. While in-plane orientations remain widely distributed without global alignment, the out-of-plane orientation landscape is biased and relatively stable. One possible origin of the distinct out-of-plane and in-plane behavior is the geometric and chemical anisotropy introduced by the planar liquid-cell window. However, because the $SiN_x$ window is amorphous, it does not provide an equivalent crystallographic reference direction within the imaging plane. Another possible explanation is the nanocrystal–window interactions governing the out-of-plane, therefore beam-parallel crystallographic direction, selected at the earliest resolved stage of growth while leaving the azimuthal rotation broadly distributed. Wang et al. showed that uniform $SiN_x$–water interfaces contain chemically heterogeneous nanoscale domains that produce spatially varying barriers for heterogeneous nanoparticle nucleation, demonstrating that the liquid-cell window can actively mediate nucleation rather than serving as an inert boundary [32]. Although the study does not include the explicit measurement of crystallographic orientation, it supports the possibility that local window chemistry influences the initial orientation distribution.

Once established, the initial out-of-plane orientation may become kinetically preserved as particle–window adhesion and confinement suppress whole-particle tilting and rotation. Chee et al. directly observed coupled translational and rotational motion of anisotropic Au nanoparticles adsorbed on $SiN_x$, showing that their motion was governed by intermittent pinning and release at specific adsorption sites; some particles exhibited only small rotations and no measurable translation.[58] Such pinning, reinforced by the confinement within the approximately 80 nm liquid channel, could inhibit changes in the out-of-plane orientation while allowing different particles to retain unrelated in-plane azimuths. We also note that varying the liquid channel thickness could establish the extent to which confinement contributes to the pinning behavior. Local reorientation can nevertheless occur during particle–particle coalescence, consistent with the abrupt orientation changes observed in the particle-resolved trajectories in Figure 2, without generating ensemble-wide alignment as each attachment event occurs with a different local collision geometry and neighboring orientation.

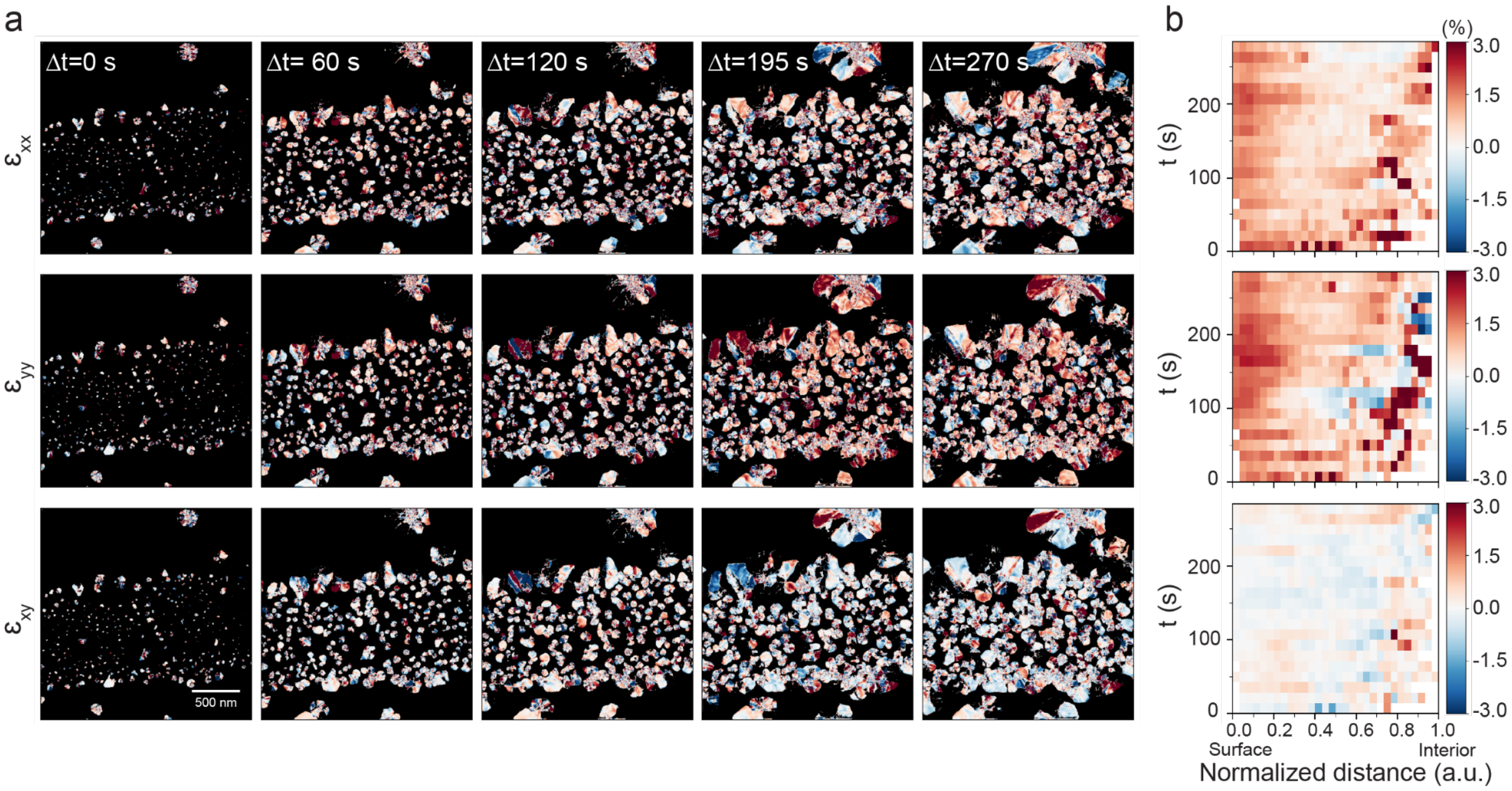

Figure 4. Temporal evolution of strain during Au nanocrystal growth. (a) Maps of strain tensor components $\epsilon_{xx}$ (top), $\epsilon_{yy}$ (middle), and $\epsilon_{xy}$ (bottom) at selected time points. (b) Heatmaps of the mean strain components as a function of normalized distance from the crystal boundary and time.

Building on the orientation-resolved analysis in Figure 3, we next examine how lattice strain evolves during nanocrystal growth. Figure 4a shows spatial maps of the strain tensor components $\epsilon_{xx}$ , $\epsilon_{yy}$, and $\epsilon_{xy}$ at representative time points, revealing the heterogeneous distribution of strain across the nanocrystal ensemble (full time series in Supporting Movie S3). Qualitatively, the strain maps indicate a progressive development of tensile strain up to 3% in the in-plane components ($\epsilon_{xx}$ and $\epsilon_{yy}$) as growth proceeds, while the shear component ($\epsilon_{xy}$) remains comparatively distributed without a clear global bias (Supporting Figure S3, Supporting Information). To further quantify the spatial localization of strain, we analyze its dependence on the distance from the nanoparticle boundary. Figure 4b presents heatmaps of the mean strain components as a function of normalized distance (0 = surface, 1 = interior) and time. This representation reveals how strain evolves not only over time but also across space within individual particles. The boundary-referenced analysis shows that tensile strain preferentially develops near the particle surface and propagates toward the interior as growth progresses, consistent with surface-driven growth processes and the accumulation of lattice mismatch during coalescence and grain evolution.

The surface-localized strain evolution can be explained by the strain accommodation during growth and coalescence. When crystallites with finite misorientation undergo oriented attachment, the interfacial region must elastically accommodate lattice mismatch prior to dislocation formation, generating a residual strain field that extends into the surrounding lattice and persists following grain merging [59]. This mechanism links the strain evolution in Figure 4 to the abrupt orientation changes associated with coalescence in Figure 2 and is consistent with prior observations of defect generation at newly formed growth and attachment interfaces during imperfect oriented attachment [38]. While these are based on ensemble behavior between nanocrystals, the contribution from individual nanocrystals should not be underestimated. Isolated Au nanocrystals can exhibit surface stress, and the magnitude and the sign depend on the shape, curvature, adsorbates, and interaction with a supporting interface [60]. These mechanisms explain the spatial gradient from surface to interior observed in Figure 4b, and the progressive buildup of strain magnitude over time as the ensemble transitions from isolated nuclei to densely packed, interacting nanocrystals, which is consistent with a nonequilibrium combination of shape-dependent surface relaxation, nanocrystal–window or nanocrystal–liquid interfacial stress, and strain incorporated at the active growth front. In contrast, the shear component exhibits weaker spatial correlation, suggesting that it is dominated by local structural heterogeneity rather than a systematic growth-induced effect. These results show that nanocrystal growth is accompanied by the buildup of tensile strain that is spatially localized near surfaces and interfaces, linking the structural pathways identified in Figures 2 and 3 to the emergence of internal lattice distortion during growth.

The magnitude of the surface-localized tensile strain observed here is also relevant to the functional properties of the resulting nanocrystals. Strain modifies orbital overlap in transition-metal surfaces and can thereby shift the d-band and alter adsorbate binding, with lattice expansion generally shifting the d states toward the Fermi level for late transition metals.[61] Importantly, strains on the order of ~3% are already sufficient to produce measurable changes in adsorption energetics and catalytic kinetics, although the magnitude and even the optimal sign of the strain depend on the metal, surface, and reaction. For CO oxidation, Nilsson Pingel *et al.* showed that experimentally measured surface- and interface-localized strains in supported Pt nanoparticles alter predicted turnover frequencies [42], while operando studies of Au nanoparticles have directly correlated evolving tensile surface strain with CO oxidation activity [62]. In electrocatalysis, Yan, Maark, Khorshidi *et al.* demonstrated that externally imposed elastic strain systematically tunes hydrogen-evolution activity, with compression favoring Pt and Ni but tension favoring Cu [63]. Calculations further predict strain-dependent H, O, and OH adsorption on Au and other late-transition-metal surfaces and identify tensile strain as a route to tuning oxygen-reduction activity on Au [64]. Thus, the ~3% tensile strain generated during nanocrystal formation lies within a regime known to perturb surface adsorption energetics. The observed dependence of strain on growth and coalescence therefore suggests that the growth pathway itself can encode a functional electronic descriptor into the resulting nanocrystal, providing a potential route to tune catalytic properties through control of nonequilibrium synthesis.

## Conclusion

In this work, we establish 5DSTEM as a powerful approach for resolving the dynamic pathways of nanocrystal growth across length scales. By combining time-resolved diffraction with real-space imaging, we directly connect ensemble growth kinetics with particle- and grain-level structural evolution. Our results reveal that, while the overall growth behavior follows surface-reaction-limited kinetics, individual nanoparticles evolve through heterogeneous pathways involving continuous growth and discrete coalescence via oriented attachment, each with distinct orientation dynamics. At the ensemble level, a stable out-of-plane orientation landscape coexists with rotationally distributed in-plane orientations, while tensile strain progressively develops from surfaces and interfaces during growth. These findings provide a unified picture linking kinetics, crystallographic evolution, and strain development in liquid-phase nanocrystal formation. Beyond the Au model system studied here, the 5DSTEM framework provides a general approach for interrogating dynamic processes in complex materials systems. In nanoscale synthesis, the ability to simultaneously track ensemble kinetics and individual particle dynamics, including discrete coalescence events and grain-by-grain orientation evolution, provides a basis for understanding how the growth pathways govern nanocrystal size, shape, and crystallographic texture. The observation that tensile strain is spatially localized and evolves dynamically during growth suggests that synthesis conditions can be leveraged to engineer strain states. By connecting with the explicit measurement of catalytic performance, this framework can be extended to catalytic and other energy conversion systems, where mechanistic insight into structural dynamics during synthesis and operation translates into functional properties of the materials. For example, in catalysis, the strain fields that develop near nanocrystal surfaces and interfaces during growth are particularly significant because they can tune catalytic activity and selectivity. With the introduction of performance evaluation, we believe the 5DSTEM framework presented in this work can be applied to broader functional materials, including bimetallic and multimetallic nanoparticles, supported catalysts, correlating synthesis pathways with structure-dependent performance.

## Methods

### 5DSTEM Experiment

The 5DSTEM data were collected on a Thermo-Fisher probe-corrected Spectra microscope using a Dectris Arina detector. The accelerating voltage was 300 kV. The data were acquired by scanning a 512 x 512 array of probe positions and at each probe position, a diffraction pattern of 192 × 192 detector pixels was recorded, using a convergence semi-angle of 0.622 mrad and a dwell time of 50 µs per probe. A 1 mM $HAuCl_4$ aqueous solution was introduced into a liquid-cell holder (Insight Chips, flow holder) with an 80 nm-thick channel. The electron dose rate of 1.14 e-/ Å$^2$ was calculated from the measured screen current and the scanned field of view.

### Contour Processing

Nanoparticle contours were extracted from each frame using automated segmentation based on the Segment Anything Model (SAM) [65], followed by centroid-based tracking across time. Individual particles were initialized from the smallest segmented objects in the first frame and matched across frames using a cost function that combines centroid proximity and area similarity. The projected area of each particle was computed directly from the segmentation mask and converted to physical units using calibrated pixel size. Contours were extracted from binary masks and aligned to their instantaneous centroid to visualize shape evolution independent of translational motion. Each particle-area trajectory was fitted using piecewise linear regression with two or three segments, requiring at least two time points per segment and sharing boundary points between adjacent segments.

**5DSTEM data analysis**

The 5DSTEM dataset was analyzed based on our previous unsupervised clustering approach [54] and the automated crystal orientation mapping (ACOM) framework [55,56], which are part of py4DSTEM (https://github.com/py4dstem/py4DSTEM.git). Face-centered cubic Au with a lattice constant of 4.08 Å was used for the template matching, and the calibrated pixel size of the reciprocal space was 0.018 $Å^{-1}$.

- **Crystal orientation quantification**

For the in-plane analysis, the crystallographic direction aligned with the image x-axis and a family-specific reference crystallographic direction were projected onto the plane perpendicular to the local beam-aligned zone axis. The signed azimuthal angle between the projected directions was then calculated about the zone axis. Pixels were assigned to the nearest out-of-plane zone-axis family (⟨001⟩, ⟨011⟩, or ⟨111⟩) based on angular proximity, and only pixels within 15° of the corresponding family were retained. The resulting azimuthal angles were reduced according to the rotational symmetry of each zone-axis family, mapping crystallographically equivalent orientations onto a common angular interval. The mean and standard deviation of the symmetry-reduced azimuthal angles were then calculated for each family at each time frame. For the out-of-plane analysis, the beam-aligned crystallographic direction was extracted at each valid pixel from the symmetry-reduced orientation representation. The angular distance between this direction and the ⟨001⟩, ⟨011⟩, and ⟨111⟩ reference zone-axis families was calculated from the dot product between the corresponding normalized direction vectors. Each pixel was assigned to the family with the smallest angular distance if that distance was ≤15°; pixels outside this cutoff were left unclassified. The fraction associated with each family was calculated relative to the total number of valid indexed pixels.

- **Strain mapping**

For the strain mapping, a transformation matrix was determined by comparing the best-fit local ACOM lattice vectors with those of a reference lattice at each probe position. The infinitesimal strain tensor was then calculated from the symmetric component of the corresponding real-space deformation matrix. For each time frame, a binary crystal mask is obtained from the valid strain region and refined using Gaussian smoothing and hole filling. The Euclidean distance transform is then computed within the mask to determine the inward distance of each pixel from the nearest boundary. This distance is normalized by the maximum inward distance in each frame, yielding a dimensionless coordinate ranging from 0 (boundary) to 1 (deep interior). For each strain component ($\epsilon_{xx}$, $\epsilon_{yy}$, and $\epsilon_{xy}$), pixels are grouped into bins based on their normalized boundary distance, and the mean strain within each bin is computed.

**Acknowledgments**

S.L. and J.A.D. acknowledge the support from the Office of Basic Energy Sciences, U.S. Department of Energy, Division of Materials Science and Engineering (DE-AC02-76SF00515). S.L. and J.A.D also acknowledge the financial support from the U.S. Department of Energy, Office of Science, National Quantum Information Science Research Centers as part of the Q-NEXT center. Additionally, S.L. and J.A.D. acknowledge financial support from the National Research Foundation of Korea (NRF) grant funded by the Korean Government (Ministry of Science and ICT) (No. RS-2024-00421181). In addition, S.L. and J.A.D. acknowledge the use and support of the Stanford Nano Shared Facilities (SNSF), supported by the National Science Foundation under award ECCS-2026822. C.O. acknowledges financial support from the U.S. Department of Energy under Contract No. DE-AC02-76SF00515 through the Basic Energy Sciences (BES) Microelectronics ESTEEM Program (101256). The authors thank Cedric Lim and Dr. Stephanie Ribet for the helpful discussions.

**Conflicts of Interest**
The authors declare no conflicts of interest.

**Data Availability Statement**
The 5DSTEM dataset of the Au nanoparticle in the liquid cell TEM is available online (https://doi.org/10.5061/dryad.1rn8pk1bk). The uploaded dataset includes a prefiltered dataset compressed in npz format and the resulting time-series orientation and strain maps in Python pickle format. The maps were generated after prefiltering and unsupervised clustering of the 4DSTEM data using the workflow described in the py4DSTEM repository (https://github.com/py4dstem/py4DSTEM.git), and the example Jupyter Notebook in the py4DSTEM_tutorials repository (https://github.com/py4dstem/py4DSTEM_tutorials). The workflow for the quantification of orientation and strain is included as part of the Supporting Information.

**Supporting Information:**

**5DSTEM of Liquid-Phase Nanocrystal Growth Bridges Ensemble Kinetics and Nanoscale Dynamics**

Authors: Serin Lee[1,*], Colin Ophus[1,*], Jennifer Dionne[1,2,*]
1: Department of Materials Science and Engineering, Stanford University, Stanford, CA, USA.
2: Department of Radiology, Stanford University School of Medicine, Stanford, CA, USA.
Corresponding authors (*): serinl@stanford.edu, cophus@stanford.edu, jdionne@stanford.edu

Supporting Movie S1: Time series of virtual annular dark-field (ADF) images.

Supporting Movie S2: Time series of crystallographic orientation maps, with in-plane orientation (left) and out-of-plane orientation (right).

Supporting Movie S3 (Strain): Time series of strain maps, including $\epsilon_{xx}$(left), $\epsilon_{yy}$(middle), and $\epsilon_{xy}$(right).

All movies are displayed at 5 frames per second (fps) and represent time-resolved datasets acquired during the Au nanocrystal growth. Each movie corresponds to the same field of view and time sequence.

Supporting Code: Jupyter notebook file for quantification of orientation and strain.

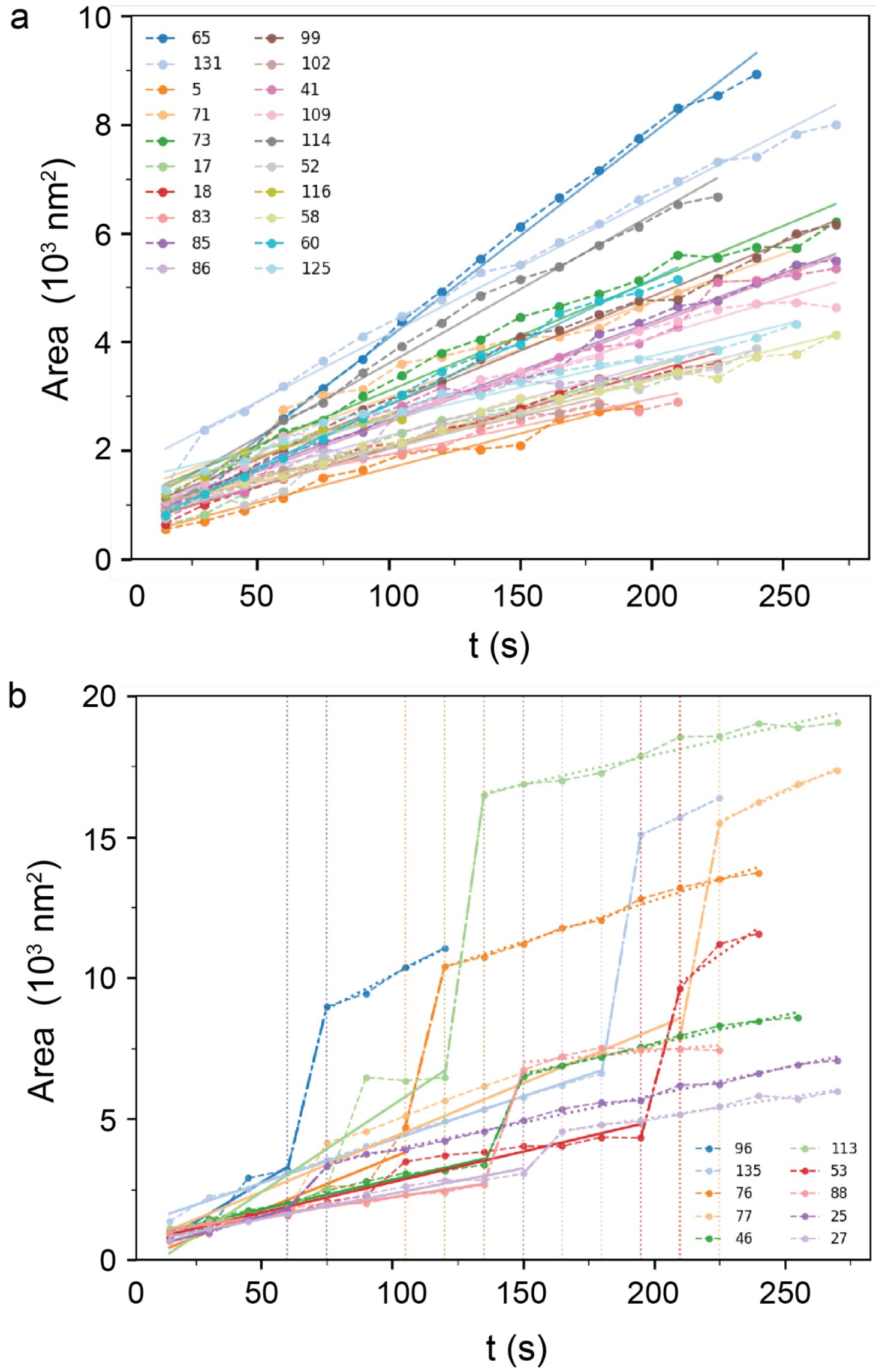

Supporting Figure S1. Linear fitting of the plots shown in Figure 2a and 2b of the main text, where the vertical lines in Figure S1b show the time of stepwise increase. A total of 150 particles were tracked, and the particles with persistent tracking continuity are shown.

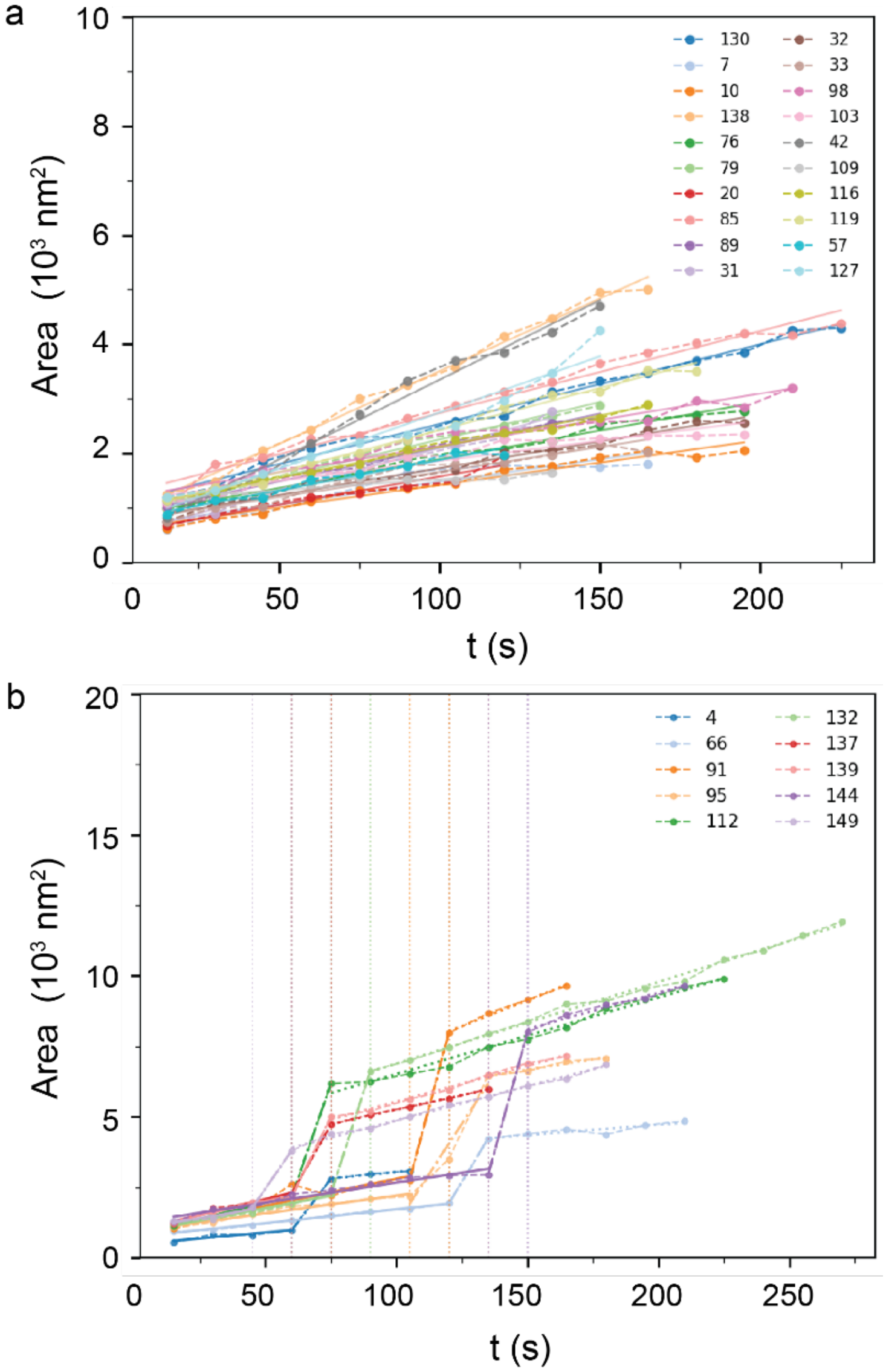

Supporting Figure S2. Linear fitting of the plots from an independent experiment under the same experimental conditions, where the vertical lines in Figure S2b show the time of stepwise increase.

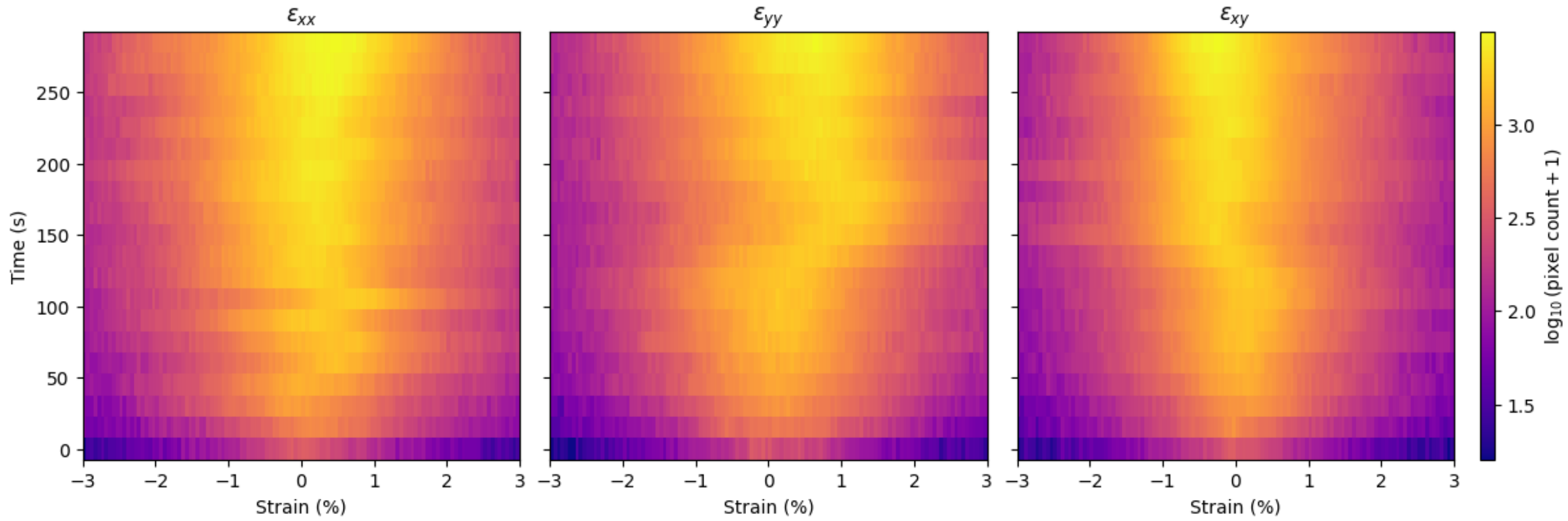

Supporting Figure S3. Heatmaps showing the time-dependent distribution of the strain components $\epsilon_{xx}$, $\epsilon_{yy}$, and $\epsilon_{xy}$. The horizontal axis represents strain (%) and the vertical axis represents time (s), with each row corresponding to a single time frame. Color indicates the logarithm of pixel counts ($\log_{10}(N+1)$) within each strain bin, highlighting the evolution of the strain distribution across the field of view.

| ID | 65 | 131 | 5 | 71 | 73 | 17 | 18 | 83 | 85 | 86 |
|---|---|---|---|---|---|---|---|---|---|---|
| $R^2$ | 0.9962 | 0.9863 | 0.9752 | 0.9183 | 0.9732 | 0.9746 | 0.9866 | 0.9746 | 0.9946 | 0.9025 |
| ID | 99 | 102 | 41 | 109 | 114 | 52 | 116 | 58 | 60 | 125 |
| $R^2$ | 0.9932 | 0.9745 | 0.9803 | 0.9749 | 0.9912 | 0.9741 | 0.9426 | 0.9782 | 0.9941 | 0.9660 |

Supporting Table 1. $R^2$ values for Figure 2a.

| ID | Segment 1 | | Segment 2 | | Segment 3 | |
|---|---|---|---|---|---|---|
| | $R^2$ | $a_1$ (slope, $nm^2/s$) | $R^2$ | $a_2$ (slope, $nm^2/s$) | $R^2$ | $a_3$ (slope, $nm^2/s$) |
| 96 | 0.8678 | 55.50 | 1.000 | 386.67 | 0.9867 | 47.67 |
| 135 | 0.9952 | 30.79 | 1.000 | 565.00 | 0.9995 | 43.33 |
| 76 | 0.8383 | 37.50 | 1.000 | 380.00 | 0.9888 | 29.53 |
| 77 | 0.9447 | 38.52 | 1.000 | 505.00 | 0.9886 | 41.67 |
| 46 | 0.9662 | 21.14 | 1.000 | 208.33 | 0.9823 | 20.95 |
| 113 | 0.8304 | 61.59 | 1.000 | 668.33 | 0.9410 | 20.96 |
| 53 | 0.9388 | 21.71 | 1.000 | 353.33 | 0.8879 | 65.00 |
| 88 | 0.9728 | 13.47 | 1.000 | 270.00 | 0.5439 | 8.00 |
| 25 | 0.9906 | 25.67 | 1.000 | 100.00 | 0.9922 | 19.43 |
| 27 | 0.9753 | 17.93 | 1.000 | 100.00 | 0.9616 | 13.91 |

Supporting Table 2. $R^2$ and slope values for Figure 2b.